\documentclass[10pt]{article}
\usepackage{amsmath}
\usepackage{amssymb}
\usepackage{graphicx}

\begin{document}

\setcounter{figure}{0}
\setcounter{table}{0}
\setcounter{footnote}{0}
\setcounter{equation}{0}

\vspace*{0.5cm}

\noindent {\Large POST-POST-NEWTONIAN LIGHT PROPAGATION WITHOUT \\
INTEGRATING THE GEODESIC EQUATIONS}
\vspace*{0.7cm}

\noindent\hspace*{1.5cm} P. TEYSSANDIER, \\
\noindent\hspace*{1.5cm} SYRTE, Observatoire de Paris, CNRS UMR8630, UPMC,\\
\noindent\hspace*{1.5cm} 61 avenue de l'Observatoire, F-75014 Paris, France\\
\noindent\hspace*{1.5cm} e-mail: Pierre.Teyssandier@obspm.fr\\

\vspace*{0.5cm}

\noindent {\large ABSTRACT.} A new derivation of the propagation direction of light is given for a 3-parameter family of static, spherically symmetric space-times within the post-post-Newtonian framework. The  emitter and the observer are both located at a finite distance. The case of a ray emitted at infinity is also treated.  

\vspace*{1cm}

\noindent {\large 1. INTRODUCTION}

\smallskip

The aim of this work is to present a new calculation of the propagation direction of light rays in a 3-parameter family of static, spherically symmetric space-times within the post-post-Newtonian framework. Rather than deriving the results from an integration of the geodesic equations, we obtain the desired expressions by a straightforward differentiation of the time delay function (see, e.
g., Teyssandier \& Le Poncin-Lafitte 2008 and Refs. therein). This study is motivated by the fact that any in-depth discussion of the highest accuracy tests of gravitational theories requires to evaluate the corrections of order higher than one in powers of the Schwarzschild radius (see, e.g., Ashby \& Bertotti 2010 for the Cassini experiment). Even for the Gaia mission, a discrepancy between the analytical post-Newtonian solution and a computational estimate has recently necessitated a thorough analysis of the post-post-Newtonian propagation of light (see Klioner \& Zschocke 2010 and Refs. therein).

\vspace*{0.7cm}

\noindent {\large 2. LIGHT DIRECTION IN SPHERICALLY SYMMETRIC SPACE-TIMES}

\smallskip

The gravitational field is assumed to be generated by an isolated spherically symmetric body of mass $M$. Setting $m=GM/c^2$, the  metric is supposed to be of the form 
\begin{equation} \label{ds2}
ds^2=\left(1 - \frac{2m}{r} + 2\beta \frac{m^2}{r^2} + \cdots\right)\left(dx^{0}\right)^2 - \left(1 + 2 \gamma \frac{m}{r} + \frac{3}{2}\epsilon \frac{m^2}{r^2} + \cdots\right)\delta_{ij}dx^{i} dx^{j},  
\end{equation}
where $r=\sqrt{\delta_{ij}x^ix^j}$, $\beta$ and $\gamma$ are the usual post-Newtonian parameters, and $\epsilon$ is a post-post-Newtonian parameter ($\beta = \gamma = \epsilon = 1$ in general relativity). We put $x^0 = ct$ and $\boldsymbol{x} = (x^i)$, with $i=1, 2, 3$.

Consider a photon emitted at a point $\boldsymbol{x}_{\scriptscriptstyle A}$ at an instant $t_{\scriptscriptstyle A}$ and received at a point $\boldsymbol{x}_{\scriptscriptstyle B}$ at an instant $t_{\scriptscriptstyle B}$. The propagation direction of this photon at any point $x$ of its path is characterized by the triple
\begin{equation} \label{tl}
\widehat{\underline{\boldsymbol{l}}}=\left(l_i /l_0\right)=\left(l_1 /l_0, l_2 /l_0, l_3 /l_0\right),  
\end{equation}
where $l_0$ and $l_i$ are the covariant components of the vector tangent to the ray, i.e. the quantities defined by $l_{\alpha}=g_{\alpha\beta}dx^{\beta}/d\lambda$, $g_{\alpha\beta}$ denoting the metric components and $\lambda$ an arbitrary parameter along the ray.

Denote by $\widehat{\underline{\boldsymbol{l}}}_{\scriptscriptstyle A}$ and $\widehat{\underline{\boldsymbol{l}}}_{\scriptscriptstyle B}$ the expressions of $\widehat{\underline{\boldsymbol{l}}}$ at points $\boldsymbol{x}_{\scriptscriptstyle A}$ and $\boldsymbol{x}_{\scriptscriptstyle B}$, respectively. In any stationary space-time, these triples can be derived from the relations (see Le Poncin-Lafitte et al. 2004)
\begin{equation} \label{wl}
\left(\frac{l_i}{l_0}\right)_{\!\scriptscriptstyle A}=c \frac{\partial {\cal T}(\boldsymbol{x}_{\scriptscriptstyle A}, \boldsymbol{x}_{\scriptscriptstyle B})}{\partial x^{i}_{\scriptscriptstyle A}}, \qquad   
\left(\frac{l_i}{l_0}\right)_{\!\scriptscriptstyle B}=-c \frac{\partial {\cal T}(\boldsymbol{x}_{\scriptscriptstyle A}, \boldsymbol{x}_{\scriptscriptstyle B})}{\partial x^{i}_{\scriptscriptstyle B}},
\end{equation}
where ${\cal T} (\boldsymbol{x}_{\scriptscriptstyle A},\boldsymbol{x}_{\scriptscriptstyle B})$ is the expression giving the travel time of a photon as a function of $\boldsymbol{x}_{\scriptscriptstyle A}$ and $\boldsymbol{x}_{\scriptscriptstyle B}$: 
\begin{equation} \label{4}
t_{\scriptscriptstyle B} - t_{\scriptscriptstyle A} = {\cal T} (\boldsymbol{x}_{\scriptscriptstyle A}, \boldsymbol{x}_{\scriptscriptstyle B}).
\end{equation}

For the metric (\ref{ds2}), ${\cal T} (\boldsymbol{x}_{\scriptscriptstyle A}, \boldsymbol{x}_{\scriptscriptstyle B})$ is given by (see, e.g., Teyssandier \& Le Poncin-Lafitte 2008):
\begin{eqnarray}
& &\mathcal{T}(\boldsymbol{x}_{\scriptscriptstyle A},\boldsymbol{x}_{\scriptscriptstyle B}) = \frac{\vert\boldsymbol{x}_{\scriptscriptstyle B}-\boldsymbol{x}_{\scriptscriptstyle A}\vert}{c} +\frac{(\gamma+1)m}{c}
\ln\left(\frac{r_{\!\scriptscriptstyle A}+r_{\!\scriptscriptstyle B}+\vert\boldsymbol{x}_{\scriptscriptstyle B}-\boldsymbol{x}_{\scriptscriptstyle A}\vert}{r_{\!\scriptscriptstyle A}+r_{\!\scriptscriptstyle B}-\vert\boldsymbol{x}_{\scriptscriptstyle B}-\boldsymbol{x}_{\scriptscriptstyle A}\vert}\right)\nonumber \\
& &\quad \qquad \qquad \qquad \quad +m^{2}\frac{\vert\boldsymbol{x}_{\scriptscriptstyle B}-\boldsymbol{x}_{\scriptscriptstyle A}\vert}{c}\left[\kappa
\frac{\arccos (\boldsymbol{n}_{\scriptscriptstyle A}.\boldsymbol{n}_{\scriptscriptstyle B})}
{\vert\boldsymbol{x}_{\scriptscriptstyle A}\times \boldsymbol{x}_{\scriptscriptstyle B}\vert}
-\frac{(\gamma+1)^2}{r_{\!\scriptscriptstyle A} r_{\!\scriptscriptstyle B}+\boldsymbol{x}_{\scriptscriptstyle A} . \boldsymbol{x}_{\scriptscriptstyle B}}\right]+ \cdots,  \label{Tss}
\end{eqnarray}
where
\begin{equation} \label{Rnn}
\boldsymbol{n}_{\scriptscriptstyle A} = \frac{\boldsymbol{x}_{\scriptscriptstyle A}}{r_{\!\scriptscriptstyle A}}, \quad \boldsymbol{n}_{\scriptscriptstyle B} = \frac{\boldsymbol{x}_{\scriptscriptstyle B}}{r_{\!\scriptscriptstyle B}}, \quad \kappa=\frac{8-4\beta+8\gamma+3\epsilon}{4}. 
\end{equation}

Substituting for ${\cal T} (\boldsymbol{x}_{\scriptscriptstyle A}, \boldsymbol{x}_{\scriptscriptstyle B})$ from Eq. (\ref{Tss}) into Eqs. (\ref{wl}) yields $\widehat{\underline{\boldsymbol{l}}}_{\scriptscriptstyle A}$ and  $\widehat{\underline{\boldsymbol{l}}}_{\scriptscriptstyle B}$ as linear combinations of $\boldsymbol{n}_{\scriptscriptstyle A}$ and $\boldsymbol{n}_{\scriptscriptstyle B}$. However, it is more convenient to introduce the unit vector $\boldsymbol{N}_{\!\scriptscriptstyle AB}$ defined by 
\begin{equation} \label{NAB}
\boldsymbol{N}_{\!\scriptscriptstyle AB} = \frac{\boldsymbol{x}_{\scriptscriptstyle B}-\boldsymbol{x}_{\scriptscriptstyle A}}{\vert\boldsymbol{x}_{\scriptscriptstyle B}-\boldsymbol{x}_{\scriptscriptstyle A}\vert}
\end{equation}
and the unit vector $\boldsymbol{P}_{\!\scriptscriptstyle AB}$ orthogonal to $\boldsymbol{N}_{\!\scriptscriptstyle AB}$ defined as $\boldsymbol{OH}/\vert \boldsymbol{OH}\vert$, $H$ being the orthogonal projection of the center $O$ of the mass $M$ on the straight line passing through $\boldsymbol{x}_{\scriptscriptstyle A}$ and $\boldsymbol{x}_{\scriptscriptstyle B}$, that is
\begin{equation} \label{PAB}
\boldsymbol{P}_{\!\scriptscriptstyle AB} = \boldsymbol{N}_{\!\scriptscriptstyle AB}\times\left(
\frac{\boldsymbol{n}_{\scriptscriptstyle A}\times\boldsymbol{n}_{\scriptscriptstyle B}}{\vert\boldsymbol{n}_{\scriptscriptstyle A}\times\boldsymbol{n}_{\scriptscriptstyle B}\vert}\right).
\end{equation}

Using Eqs. (\ref{Tss})-(\ref{PAB}), we deduce the following proposition from Eqs. (\ref{wl}).

$ $

{\bf Proposition 1.} {\em The triples} $\underline{\widehat{\boldsymbol{l}}}_{\scriptscriptstyle A}$ {\em and} $\underline{\widehat{\boldsymbol{l}}}_{\scriptscriptstyle B}$ {\em are given by}
\begin{eqnarray}
& &\!\!\!\!\!\underline{\widehat{\boldsymbol{l}}}_{\scriptscriptstyle A}=-\boldsymbol{N}_{\!\scriptscriptstyle AB}-\frac{m}{r_{\!\scriptscriptstyle A}}\Bigg\lbrace(\gamma + 1)+\frac{m}{r_{\!\scriptscriptstyle A}}\bigg\lbrack\kappa-\frac{(\gamma+1)^2}{1+\boldsymbol{n}_{\scriptscriptstyle A}.\boldsymbol{n}_{\scriptscriptstyle B}}\bigg\rbrack\Bigg\rbrace\boldsymbol{N}_{\!\scriptscriptstyle AB}\nonumber\\
& &\qquad\qquad\,\,-\,\frac{m}{r_{\!\scriptscriptstyle A}}\,\Bigg\lbrace(\gamma + 1)\frac{\vert\boldsymbol{n}_{\scriptscriptstyle A}\times\boldsymbol{n}_{\scriptscriptstyle B}\vert}{1+\boldsymbol{n}_{\scriptscriptstyle A}.\boldsymbol{n}_{\scriptscriptstyle B}}+\frac{m}{r_{\!\scriptscriptstyle A}}\frac{1}{\vert\boldsymbol{n}_{\scriptscriptstyle A}\times\boldsymbol{n}_{\scriptscriptstyle B}\vert}\bigg\lbrace \kappa\bigg\lbrack \frac{\arccos(\boldsymbol{n}_{\scriptscriptstyle A}.\boldsymbol{n}_{\scriptscriptstyle B})}{\vert\boldsymbol{n}_{\scriptscriptstyle A}\times\boldsymbol{n}_{\scriptscriptstyle B}\vert}\left(1-\frac{r_{\!\scriptscriptstyle A}}{r_{\!\scriptscriptstyle B}}\boldsymbol{n}_{\scriptscriptstyle A}.\boldsymbol{n}_{\scriptscriptstyle B}\right)\qquad\nonumber \\
& &\qquad\qquad\qquad\qquad\qquad\qquad\quad+\frac{r_{\!\scriptscriptstyle A}}{r_{\!\scriptscriptstyle B}}-\boldsymbol{n}_{\scriptscriptstyle A}.\boldsymbol{n}_{\scriptscriptstyle B}\bigg\rbrack -(\gamma+1)^2\left(1+\frac{r_{\!\scriptscriptstyle A}}{r_{\!\scriptscriptstyle B}}\right)\frac{1-\boldsymbol{n}_{\scriptscriptstyle A}.\boldsymbol{n}_{\scriptscriptstyle B}}{1+\boldsymbol{n}_{\scriptscriptstyle A}.\boldsymbol{n}_{\scriptscriptstyle B}}\bigg\rbrace\Bigg\rbrace\boldsymbol{P}_{\!\scriptscriptstyle AB}\!\!\!\!\!\!\label{dirA2}
\end{eqnarray}
{\em and}
\begin{eqnarray}
& &\!\!\!\!\!\underline{\widehat{\boldsymbol{l}}}_{\scriptscriptstyle B}=-\boldsymbol{N}_{\!\scriptscriptstyle AB}-\frac{m}{r_{\!\scriptscriptstyle B}}\Bigg\lbrace\gamma + 1+\frac{m}{r_{\!\scriptscriptstyle B}}\bigg\lbrack \kappa-\frac{(\gamma+1)^2}{1+\boldsymbol{n}_{\scriptscriptstyle A}.\boldsymbol{n}_{\scriptscriptstyle B}}\bigg\rbrack\Bigg\rbrace\boldsymbol{N}_{\!\scriptscriptstyle AB} \nonumber \\
& &\qquad\qquad\,\,+\,\frac{m}{r_{\!\scriptscriptstyle B}}\,\Bigg\lbrace(\gamma + 1)\frac{\vert\boldsymbol{n}_{\scriptscriptstyle A}\times\boldsymbol{n}_{\scriptscriptstyle B}\vert}{1+\boldsymbol{n}_{\scriptscriptstyle A}.\boldsymbol{n}_{\scriptscriptstyle B}}+\frac{m}{r_{\!\scriptscriptstyle B}}\frac{1}{\vert\boldsymbol{n}_{\scriptscriptstyle A}\times\boldsymbol{n}_{\scriptscriptstyle B}\vert}\bigg\lbrace \kappa\bigg\lbrack \frac{\arccos(\boldsymbol{n}_{\scriptscriptstyle A}.\boldsymbol{n}_{\scriptscriptstyle B})}{\vert\boldsymbol{n}_{\scriptscriptstyle A}\times\boldsymbol{n}_{\scriptscriptstyle B}\vert}\left(1-\frac{r_{\!\scriptscriptstyle B}}{r_{\!\scriptscriptstyle A}}\boldsymbol{n}_{\scriptscriptstyle A}.\boldsymbol{n}_{\scriptscriptstyle B}\right)\qquad\nonumber \\
& &\qquad\qquad\qquad\qquad\qquad\qquad\quad+\frac{r_{\!\scriptscriptstyle B}}{r_{\!\scriptscriptstyle A}}-\boldsymbol{n}_{\scriptscriptstyle A}.\boldsymbol{n}_{\scriptscriptstyle B}\bigg\rbrack -(\gamma+1)^2\left(1+\frac{r_{\!\scriptscriptstyle B}}{r_{\!\scriptscriptstyle A}}\right)\frac{1-\boldsymbol{n}_{\scriptscriptstyle A}.\boldsymbol{n}_{\scriptscriptstyle B}}{1+\boldsymbol{n}_{\scriptscriptstyle A}.\boldsymbol{n}_{\scriptscriptstyle B}}\bigg\rbrace\Bigg\rbrace\boldsymbol{P}_{\!\scriptscriptstyle AB},\!\!\!\!\!\! \label{dirB2}
\end{eqnarray}
{\em respectively.}

$ $  

In any static, spherically symmetric space-time the geodesic equations imply that the vector $\boldsymbol{L}$ defined as $\boldsymbol{L} = - \boldsymbol{x} \times \underline{\widehat{\boldsymbol{l}}}$ is a constant of the motion. The null geodesics considered here are assumed to be unbound. Consequently the magnitude of $\boldsymbol{L}$ is such that
$\vert\boldsymbol{L}\vert = lim_{\vert\boldsymbol{x}\vert \rightarrow \infty}\left\vert\boldsymbol{x}\times d\boldsymbol{x}/cdt\right\vert$
since $\underline{\widehat{\boldsymbol{l}}} \longrightarrow -(d\boldsymbol{x}/cdt)_{\infty}$ when $\vert\boldsymbol{x}\vert \longrightarrow \infty$. So the quantity $b$ defined by
\begin{equation} \label{imp2}
b = \vert-\boldsymbol{x}\times \underline{\widehat{\boldsymbol{l}}}\,\vert
\end{equation}
is the Euclidean distance between the asymptote to the ray and the line parallel to this asymptote passing through the center $O$ as measured by an inertial observer at rest at infinity. Hence $b$ may be considered as {\em the impact parameter} of the ray (see, e.g., Chandrasekhar 1983). Besides its geometric meaning, $b$ presents the interest to be {\em intrinsic}, since it corresponds to a quantity which could be really measured.

Substituting for $\underline{\widehat{\boldsymbol{l}}}_{\scriptscriptstyle B}$ from Eq. (\ref{dirB2}) into Eq. (\ref{imp2}), introducing the zeroth-order distance of closest approach $r_c$ defined as
\begin{equation} \label{rc1}
r_c=\frac{r_{\!\scriptscriptstyle A} r_{\!\scriptscriptstyle B}}{\vert\boldsymbol{x}_{\scriptscriptstyle B}-\boldsymbol{x}_{\scriptscriptstyle A}\vert}\vert \boldsymbol{n}_{\scriptscriptstyle A}\times\boldsymbol{n}_{\scriptscriptstyle B}\vert,
\end{equation} 
and then using $(r_{\!\scriptscriptstyle A} +r_{\!\scriptscriptstyle B})\vert \boldsymbol{n}_{\scriptscriptstyle A}\times\boldsymbol{n}_{\scriptscriptstyle B}\vert/\vert\boldsymbol{x}_{\scriptscriptstyle B}-\boldsymbol{x}_{\scriptscriptstyle A}\vert = \vert\boldsymbol{N}_{\!\scriptscriptstyle AB}\times\boldsymbol{n}_{\scriptscriptstyle A}\vert + \vert\boldsymbol{N}_{\!\scriptscriptstyle AB}\times\boldsymbol{n}_{\scriptscriptstyle B}\vert$, we get
\begin{equation} \label{imp1}
b=r_c \left[1 + 
\frac{(\gamma+1)m}{r_c}\frac{\vert\boldsymbol{N}_{\!\scriptscriptstyle AB}\times\boldsymbol{n}_{\scriptscriptstyle A}\vert + \vert\boldsymbol{N}_{\!\scriptscriptstyle AB}\times\boldsymbol{n}_{\scriptscriptstyle B}\vert}{1 + \boldsymbol{n}_{\scriptscriptstyle A}.
\boldsymbol{n}_{\scriptscriptstyle B}}+\cdots \right]. 
\end{equation} 

Using this expansion of $b$, we obtain the proposition which follows.

$ $

{\bf Proposition 2.} {\em In terms of the impact parameter $b$, the triples} $\underline{\widehat{\boldsymbol{l}}}_{\scriptscriptstyle A}$ {\em and} $\underline{\widehat{\boldsymbol{l}}}_{\scriptscriptstyle B}$ {\em may be written as}
\begin{eqnarray}
& &\!\!\!\!\!\underline{\widehat{\boldsymbol{l}}}_{\scriptscriptstyle A}=-\boldsymbol{N}_{\!\scriptscriptstyle AB} -\frac{m\vert\boldsymbol{N}_{\!\scriptscriptstyle AB}\times\boldsymbol{n}_{\scriptscriptstyle A}\vert}{b}\bigg\lbrace\gamma+1+\frac{m}{b}\bigg\lbrack \kappa\vert\boldsymbol{N}_{\!\scriptscriptstyle AB}\times\boldsymbol{n}_{\scriptscriptstyle A}\vert + (\gamma+1)^2 \frac{\vert\boldsymbol{N}_{\!\scriptscriptstyle AB}\times\boldsymbol{n}_{\scriptscriptstyle B}\vert}{1+\boldsymbol{n}_{\scriptscriptstyle A}.\boldsymbol{n}_{\scriptscriptstyle B}}\bigg\rbrack\bigg\rbrace\boldsymbol{N}_{\!\scriptscriptstyle AB} \qquad\nonumber \\
& &\qquad\qquad\quad-\frac{m\vert\boldsymbol{N}_{\!\scriptscriptstyle AB}\times\boldsymbol{n}_{\scriptscriptstyle A}\vert}{b}\bigg\lbrace(\gamma+1)\frac{\vert\boldsymbol{n}_{\scriptscriptstyle A}\times\boldsymbol{n}_{\scriptscriptstyle B}\vert}{1+\boldsymbol{n}_{\scriptscriptstyle A}.\boldsymbol{n}_{\scriptscriptstyle B}}\nonumber \\
& &\qquad\qquad\qquad\qquad\qquad\qquad\quad+\frac{\kappa m}{b}\left[\frac{\arccos(\boldsymbol{n}_{\scriptscriptstyle A}.\boldsymbol{n}_{\scriptscriptstyle B})}{\vert\boldsymbol{n}_{\scriptscriptstyle A}\times\boldsymbol{n}_{\scriptscriptstyle B}\vert}\boldsymbol{N}_{\!\scriptscriptstyle AB}.\boldsymbol{n}_{\scriptscriptstyle B}-\boldsymbol{N}_{\!\scriptscriptstyle AB}.\boldsymbol{n}_{\scriptscriptstyle A}\right]\bigg\rbrace\boldsymbol{P}_{\!\scriptscriptstyle AB}, \label{dirA2b}
\end{eqnarray}
\begin{eqnarray}
& &\!\!\!\!\!\!\underline{\widehat{\boldsymbol{l}}}_{\scriptscriptstyle B}=-\boldsymbol{N}_{\!\scriptscriptstyle AB} -\frac{m\vert\boldsymbol{N}_{\!\scriptscriptstyle AB}\times\boldsymbol{n}_{\scriptscriptstyle B}\vert}{b}\bigg\lbrace\gamma+1+\frac{m}{b}\bigg\lbrack \kappa\vert\boldsymbol{N}_{\!\scriptscriptstyle AB}\times\boldsymbol{n}_{\scriptscriptstyle B}\vert + (\gamma+1)^2 \frac{\vert\boldsymbol{N}_{\!\scriptscriptstyle AB}\times\boldsymbol{n}_{\scriptscriptstyle A}\vert}{1+\boldsymbol{n}_{\scriptscriptstyle A}.\boldsymbol{n}_{\scriptscriptstyle B}}\bigg\rbrack\bigg\rbrace\boldsymbol{N}_{\!\scriptscriptstyle AB} \qquad\nonumber \\
& &\qquad\qquad\;+\frac{m\vert\boldsymbol{N}_{\!\scriptscriptstyle AB}\times\boldsymbol{n}_{\scriptscriptstyle B}\vert}{b}\,\bigg\lbrace(\gamma+1)\frac{\vert\boldsymbol{n}_{\scriptscriptstyle A}\times\boldsymbol{n}_{\scriptscriptstyle B}\vert}{1+\boldsymbol{n}_{\scriptscriptstyle A}.\boldsymbol{n}_{\scriptscriptstyle B}}\nonumber \\
& &\qquad\qquad\qquad\qquad\qquad\qquad\;-\frac{\kappa m}{b}\left[\frac{\arccos(\boldsymbol{n}_{\scriptscriptstyle A}.\boldsymbol{n}_{\scriptscriptstyle B})}{\vert\boldsymbol{n}_{\scriptscriptstyle A}\times\boldsymbol{n}_{\scriptscriptstyle B}\vert}\boldsymbol{N}_{\!\scriptscriptstyle AB}.\boldsymbol{n}_{\scriptscriptstyle A}-\boldsymbol{N}_{\!\scriptscriptstyle AB}.\boldsymbol{n}_{\scriptscriptstyle B}\right]\bigg\rbrace\boldsymbol{P}_{\!\scriptscriptstyle AB}
.\label{dirB2b}
\end{eqnarray}

\vspace*{0.7cm}

\noindent {\large 3. DEFLECTION OF A LIGHT RAY EMITTED AT INFINITY}

\smallskip

Assume now that the ray arriving at $\boldsymbol{ x}_{\scriptscriptstyle B}$  is emitted at infinity in a direction defined by a unit vector $\boldsymbol{N}_{\!e}$. Substituting $\boldsymbol{N}_{\!e}$ for $\boldsymbol{N}_{\!\scriptscriptstyle AB}$ and $-\boldsymbol{N}_{\!e}$ for $\boldsymbol{n}_{\!\scriptscriptstyle A}$ in Eq. (\ref{dirB2b}) yields the expression of ${\widehat{\boldsymbol{l}}}_{\scriptscriptstyle B}$, where $b$ is furnished by the limit of Eqs. (\ref{rc1}) and (\ref{imp1}) when $r_{\!\scriptscriptstyle A}\rightarrow\infty$ and $\boldsymbol{n}_{\!\scriptscriptstyle A}\rightarrow-\boldsymbol{N}_{\!e}$. We can set a proposition as follows.

$ $

{\bf Proposition 3.} {\em For a light ray emitted at infinity in a direction} $\boldsymbol{N}_{\!e}$ {\em and arriving at} $\boldsymbol{x}_{\scriptscriptstyle B}$, $\underline{\widehat{\boldsymbol{l}}}_{\scriptscriptstyle B}$ {\em is given by} 
\begin{eqnarray}
& &\!\!\!\underline{\widehat{\boldsymbol{l}}}_{\scriptscriptstyle B}=-\boldsymbol{N}_{\!e}-\frac{m\vert\boldsymbol{N}_{\!e}\times\boldsymbol{n}_{\scriptscriptstyle B}\vert}{b}\left[\gamma+1+\frac{\kappa m\vert\boldsymbol{N}_{\!e}\times\boldsymbol{n}_{\scriptscriptstyle B}\vert}{b}\right]\boldsymbol{N}_{\!e}\qquad\qquad\qquad\qquad\nonumber \\
& &\qquad\qquad\quad+\frac{m}{b}\bigg\lbrace(\gamma+1)(1+\boldsymbol{N}_{\!e}.\boldsymbol{n}_{\scriptscriptstyle B})+\frac{\kappa m}{b}\left[\pi-\arccos(\boldsymbol{N}_{\!e}.\boldsymbol{n}_{\scriptscriptstyle B})\qquad\qquad\qquad\right.\nonumber \\
& &\left.\qquad\qquad\qquad\qquad\qquad\qquad\qquad\qquad\qquad\qquad+\vert\boldsymbol{N}_{\!e}\times\boldsymbol{n}_{\scriptscriptstyle B}\vert \boldsymbol{N}_{\!e}.\boldsymbol{n}_{\scriptscriptstyle B}\right]\bigg\rbrace
\boldsymbol{P}_{\!\scriptscriptstyle B}(\boldsymbol{N}_{\!e}),\qquad \quad \label{dirB3}
\end{eqnarray}
{\em where} $\boldsymbol{P}_{\!\scriptscriptstyle B}(\boldsymbol{N}_{\!e})$ {\em is the unit vector orthogonal to} $\boldsymbol{N}_{\!e}$ {\em defined as}
\begin{equation} \label{PAB2}
\boldsymbol{P}_{\!\scriptscriptstyle B}(\boldsymbol{N}_{\!e})=-\boldsymbol{N}_{\!e}\times\frac{\boldsymbol{N}_{\!e}\times\boldsymbol{n}_{\scriptscriptstyle B}}{\vert\boldsymbol{N}_{\!e}\times\boldsymbol{n}_{\scriptscriptstyle B}\vert} 
\end{equation}
{\em and} $b$ {\em is the impact parameter of the ray, namely}
\begin{equation} \label{imp3}
b=r_{c}\left[1+\frac{(\gamma+1)m}{r_{c}}\frac{\vert \boldsymbol{N}_{\!e}\times\boldsymbol{n}_{\scriptscriptstyle B}\vert}{1-\boldsymbol{N}_{\!e}.\boldsymbol{n}_{\scriptscriptstyle B}}+\cdots\right],
\end{equation}
{\em with} $r_{c} = r_{\!\scriptscriptstyle B}\vert \boldsymbol{N}_{\!e}\times\boldsymbol{n}_{\scriptscriptstyle B}\vert$.

$ $

The deflection of the ray at point $\boldsymbol{x}_{\scriptscriptstyle B}$ may be characterized by the angle $\Delta\chi_{\scriptscriptstyle B}$ made by the vector $\boldsymbol{N}_{\!e}$ and a vector tangent to the ray at $\boldsymbol{x}_{\scriptscriptstyle B}$. We have
\begin{equation} \label{chiB1}
\Delta\chi_{\scriptscriptstyle B}=\frac{\vert\boldsymbol{N}_{\!e}\times\underline{\widehat{\boldsymbol{l}}}_{\scriptscriptstyle B}\vert}{\vert\underline{\widehat{\boldsymbol{l}}}_{\scriptscriptstyle B}\vert}+O(1/c^6).
\end{equation}

Substituting for $\underline{\widehat{\boldsymbol{l}}}_{\scriptscriptstyle B}$ from Eq. (\ref{dirB3}) into Eq. (\ref{chiB1}), and then introducing the angle $\phi_{\scriptscriptstyle B}$ between $\boldsymbol{N}_{\!e}$ and $\boldsymbol{n}_{\scriptscriptstyle B}$ defined by 
\begin{equation} \label{phiB}
\boldsymbol{N}_{\!e}.\boldsymbol{n}_{\scriptscriptstyle B}=\cos\phi_{\scriptscriptstyle B},\qquad  0\leq\phi_{\scriptscriptstyle B}\leq\pi,
\end{equation}
we get
\begin{equation} \label{chiB2}
\Delta\chi_{\scriptscriptstyle B}=\frac{(\gamma+1)GM}{c^2 b}(1\!+\!\cos\phi_{\scriptscriptstyle B})+\frac{G^2\!M^2}{c^4 b^2}\bigg\lbrack\kappa\!\left(\!\pi\!-\!\phi_{\scriptscriptstyle B}\!+\!\frac{1}{2}\sin2\phi_{\scriptscriptstyle B}\right)\!\!-\!(\gamma+1)^2(1\!+\cos\phi_{\scriptscriptstyle B}\!)\sin\phi_{\scriptscriptstyle B}\bigg\rbrack, 
\end{equation}
where the impact parameter given by Eq. (\ref{imp3}) may be rewritten as
\begin{equation} \label{impch}
b=r_{c}\bigg\lbrack 1+\frac{(\gamma+1)GM}{c^2 r_{c}}\frac{\sin\phi_{\scriptscriptstyle B}}{1-\cos\phi_{\scriptscriptstyle B}}+\cdots\bigg\rbrack, \qquad r_c=r_{\!\scriptscriptstyle B}\sin\phi_{\scriptscriptstyle B}.
\end{equation}

It may be seen from the formulas
 given in Teyssandier \& Le Poncin-Lafitte 2006 that $\phi_{\scriptscriptstyle B}+\Delta\chi_{\scriptscriptstyle B}$ is the angular distance between the center $O$ and the source at infinity as measured at $\boldsymbol{x}_{\scriptscriptstyle B}$ by a static observer, i.e. an observer at rest with respect to the coordinates $x^i$. It will be shown in a subsequent paper that this property implies that $\Delta\chi_{\scriptscriptstyle B}$ can be regarded as an {\em intrinsic} quantity.

The $1/c^2$ term in Eq. (\ref{chiB2}) is currently used in VLBI astrometry. If $b$ is replaced by its coordinate expression (\ref{impch}), it may be seen that $\Delta\chi_{\scriptscriptstyle B}$ is given by an expression as follows
\begin{eqnarray}
& &\Delta\chi_{\scriptscriptstyle B}=\frac{(\gamma+1)GM}{c^2 r_c}(1\!+\!\cos\phi_{\scriptscriptstyle B})+\frac{G^2\!M^2}{c^4 r_c^2}\bigg\lbrack\kappa\!\left(\!\pi\!-\!\phi_{\scriptscriptstyle B}\!+\!\frac{1}{2}\sin2\phi_{\scriptscriptstyle B}\right)\!\!-\!(\gamma+1)^2(1\!+\cos\phi_{\scriptscriptstyle B}\!)\sin\phi_{\scriptscriptstyle B}\qquad \qquad\nonumber \\
& &\qquad\qquad\qquad\qquad\qquad\qquad\qquad\qquad\qquad\underbrace{-(\gamma+1)^2\frac{(1+\cos\phi_{\scriptscriptstyle B})^2}{\sin\phi_{\scriptscriptstyle B}}}\bigg\rbrack. \label{chiB3}
\end{eqnarray}  

For a ray grazing a mass $M$ of radius $r_{0}$, the underbraced term in the r.h.s. of Eq. (\ref{chiB3}) generates a post-post-Newtonian contribution $(\Delta\chi^{\scriptscriptstyle (2)}_{\scriptscriptstyle B})_{grazing} \approx- 4 (\gamma + 1)^2 (GM/c^2r_{0})^2 (r_{\scriptscriptstyle B}/r_{0})$ which can be great if $r_{\scriptscriptstyle B}\gg r_{0}$. For Jupiter, $(\Delta\chi^{\scriptscriptstyle (2)}_{\scriptscriptstyle B})_{grazing}= 16.1 \, \mu \mbox{as}$ if the observer is located at a distance from Jupiter $r_{\scriptscriptstyle B}=6$ AU: this value is appreciably greater than the level of accuracy expected for Gaia. However, this `enhanced' term is due to the use of the coordinate-dependent quantity $r_{c}$ instead of the intrinsic impact parameter $b$. This result confirms the conclusion recently drawn in Klioner \& Zschocke 2010.

\vspace*{0.7cm}

\noindent {\large 4. CONCLUSION}

\smallskip

Deriving the second-order terms in the propagation direction of light from the time transfer function rather than from the null geodesic equations is a very elegant and powerful procedure. The application of this method to a ray emitted at infinity and received by a static observer located at a finite distance from the central mass is easy and yields an intrinsic characterization of the gravitational bending of light.

\vspace*{0.7cm}

\noindent {\large 5. REFERENCES}
% Please type the reference as follows
% Name Initial, year, "title", journal, vol. , pp. x-x.
%
% Examples:
%
% Author1, N., Author2, N., 2000, ``Title of the paper'', 
% \aa 111, pp. 111--222.
%
% Author2, N., Author3, N., 2003, ``Title of the paper'',
% \jgr (Solid Earth), 111(B5), doi: 10.1000/2002JB001111.
%
% PLEASE DO NOT USE ANY SPECIAL FONTS 
% (no italics, no boldface, etc.)
%
{

\leftskip=5mm
\parindent=-5mm

\smallskip

Ashby, N., Bertotti, B.,  2010, Class. Quantum Grav. 27, 145013 (27pp).

Chandrasekhar, S., 1983, ``The Mathematical Theory of Black Holes", Clarendon Press.

Klioner, S. A., Zschocke, S., 2010, Class. Quantum Grav.  27, 075015 (25pp).

Le Poncin-Lafitte, C., Linet, B., Teyssandier, P., 2004, Class. Quantum Grav. 21, 4463 (20pp).

Teyssandier, P., Le Poncin-Lafitte, C., 2006, arXiv:gr-qc/0611078v1.

Teyssandier, P., Le Poncin-Lafitte, C., 2008, Class. Quantum Grav. 25, 145020 (12pp).

}

\end{document}